\documentclass[runningheads]{llncs}
\usepackage[T1]{fontenc}
\usepackage{graphicx}
\usepackage[utf8]{inputenc}
\usepackage{times}
\usepackage{amsmath,amssymb}
\usepackage{booktabs}
\usepackage{hyperref}
\usepackage{enumitem}
\usepackage{algorithm}
\usepackage{algorithmic}
\usepackage{microtype}
\usepackage{multirow}
\usepackage{subcaption}
\usepackage{tabularx}
\usepackage{xcolor}
\usepackage{array}
\usepackage{colortbl}

\usepackage{amsfonts}
\usepackage{lipsum}
\usepackage[numbers,sort]{natbib}
\definecolor{persuadercolor}{RGB}{220,235,250}  
\definecolor{persuadeecolor}{RGB}{255,242,204} 

\usepackage{wrapfig}

\title{Personality-Aware Reinforcement Learning for Persuasive Dialogue with LLM-Driven Simulation}
\author{Donghuo Zeng \orcidID{0000-0002-6425-6270
} \and
Roberto Legaspi \orcidID{0000-0001-8909-635X} \and
Kazushi Ikeda \orcidID{0009-0000-9563-760X}}
\authorrunning{Zeng et al.}
\institute{KDDI Research, Inc., Saitama, Japan \\ \email{\{do-zeng, ro-legaspi, kz-ikeda\}@kddi-research.jp}}

\begin{document}
\maketitle

\begin{abstract}
Effective persuasive dialogue agents adapt their strategies to individual users, accounting for the evolution of their psychological states and intentions throughout conversations. We present a personality-aware reinforcement learning approach comprised of three main modules: (1) a Strategy-Oriented Interaction Framework, which serves as an agenda-based strategy controller that selects strategy-level actions and generate responses via Maximal Marginal Relevance (MMR) retrieval to ensure contextual relevance, diversity, and scalable data generation; (2) Personality-Aware User Representation Learning, which produces an 81-dimensional mixed-type embedding predicted at each turn from recent exchanges and appended to the reinforcement learning state; and (3) a Dueling Double DQN (D3QN) model and Reward Prediction, in which the policy is conditioned on dialogue history and turn-level personality estimates and trained using a composite reward incorporating agreement intent, donation amount, and change-of-mind penalties. We use an agenda-based LLM simulation pipeline to generate diverse interactions, from which personality estimation is inferred from the generated utterances. Experiments on the PersuasionForGood (P4G) dataset augmented with simulated dialogues reveal three main findings: (i) turn-level personality conditioning improves policy adaptability and cumulative persuasion rewards; (ii) LLM-driven simulation enhances generalization to unseen user behaviors; and (iii) incorporating a change-of-mind penalty reduces post-agreement retractions while slightly improving donation outcomes. These results demonstrate that structured interaction, dynamic personality estimation, and behaviorally informed rewards together yield more effective persuasive policies.
\end{abstract}

\section{Introduction}
Persuasive dialogue systems~\cite{prakken2006formal,torning2009persuasive,yoshino2018dialogue} aim to influence users’ beliefs, behaviors, and decision-making through multi-turn interactions. They have shown promise in applications such as charitable fundraising~\cite{wang2019persuasion, zeng2024counterfactual, zeng2025generative, zeng2025causal}, health promotion~\cite{torkamaan2021integrating, liu2022persuasive}, and marketing~\cite{braca2023developing}. While recent large language models (LLMs), including ChatGPT, Gemini, and Grok, can generate fluent and contextually appropriate responses, they often lack stable behavioral grounding. In particular, generative responses may overlook subtle psychological cues, such as user personality or prior intent, which are crucial for persuasion. These limitations arise because LLMs, trained for general conversational quality, are not explicitly optimized to model individualized, long-term behavioral change or strategic dialogue planning~\cite{park2023socialsim, ouyang2022training}. 

In practice, two limitations hinder progress. First, static persona models miss dynamic changes. Most reinforcement learning (RL) based dialogue systems either ignore user individuality or model it through static persona profiles~\cite{li2016persona, zhou2023predictpersona}. However, persuasive interactions are inherently dynamic, i.e., a user’s psychological state and intentions evolve throughout the conversation. Static modeling fails to capture moment-to-moment transitions, leading to suboptimal strategy selection. 
Second, simulation and data coverage are insufficient constrain the training of robust policies. Large-scale RL training requires diverse, realistic user feedback, yet the annotated persuasive datasets like PersuasionForGood (\textbf{P4G})~\cite{wang2019persuasion} are expensive to produce and limited in coverage. Furthermore, traditional rule-based or template simulators cannot emulate nuanced user behaviors or personality-driven reactions~\cite{shi2022user}. Although LLMs offer a potential solution, they must be governed by structured interaction frameworks to ensure behavioral diversity, precise control, and ethical reliability. 

To address these challenges, we propose a personality-aware reinforcement learning architecture for persuasive dialogue that combines structured interaction design, dynamic user modeling, and composite reward optimization. Our system consists of three key components:
\begin{enumerate}
    \item \textit{Strategy-Oriented Interaction Framework}. We design an agenda-based framework in which the system (persuader) selects strategy-level actions and realizes them as utterances using Maximal Marginal Relevance (MMR) to ensure contextual relevance and diversity. The user (persuadee) is simulated through an LLM-driven agenda-based prompting scheme that produces diverse, realistic responses. Rather than assuming the LLM maintains a fixed persona, we leverage its generated utterances and extract persona-consistent signals to obtain turn-level personality estimates used by the policy. This framework enables controlled dataset expansion and exposes the agent to realistic trajectories and previously unseen behaviors beyond the original \textbf{P4G} dataset.
    \item \textit{Personality-Aware User Representation}. Each user is represented by an 81-dimensional embedding derived from both continuous and categorical personality traits, dynamically inferred at each turn. The inferred turn-level personality is concatenated with dialogue history to form the RL state, enabling the agent to condition action selection on the user’s current observable "persona trajectory".
    \item \textit{Reinforcement Learning with D3QN and Reward Prediction}. We employ a D3QN model to optimize strategy selection. The model is trained on a composite reward function encompassing short-term agreement intent, donation amount, and a novel change-of-mind penalty designed to discourage fickle commitments or retractions. 
\end{enumerate}

We leverage the \textbf{P4G} dataset to train our \textit{Strategy-Oriented Interaction Framework}, which subsequently serves as a simulation environment for generating the diverse training and evaluation trajectories required for our personality-aware RL architecture. Empirical evaluations reveal three primary findings: (1) conditioning the policy on turn-level personality consistently increases cumulative persuasion rewards; (2) LLM-driven simulation enhances generalization to previously unseen user behaviors; and (3) adding a change-of-mind reward term reduces the frequency and magnitude of post-agreement retractions while slightly improving donations. Experimental results demonstrate that the synergy of structured interaction, dynamic personality estimation, and behaviorally informed rewards yields more effective persuasive policies.
%-----------------------------------------------------------------------------------------------------------------
%------------------------------------------Related work------------------------------------------------------
%--------------------------------------------------------------------------------------------------------------------
\section{Related Work}
Our work builds on four research areas: persuasive dialogue, personalization and personality modeling, reinforcement learning for dialogue, and LLM-based simulation.

\subsection{Persuasive Dialogue} Agenda-based and strategy-driven approaches have long been used to structure persuasive interactions by encoding goals and permissible strategy transitions. These frameworks enable controlled policy learning and interpretable behavior~\cite{wang2019persuasion, kawano2022multimodal}. Recent multimodal persuasive corpora further demonstrate the importance of context and interpersonal cues in persuasion tasks \cite{kawano2022multimodal}. However, most prior systems rely on static strategies or handcrafted rules, limiting adaptability to dynamic user responses.

\subsection{Personalization and Personality Modeling} Personalized dialogue systems aim to tailor responses according to user characteristics, such as persona or psychological traits. Early studies modeled users via static profiles or explicit persona descriptions \cite{cho2022personalized, wu2023personaattention}, while later work sought to infer latent personality representations directly from dialogue context \cite{zhou2023predictpersona, tang2023contrastive}. Recent advances also emphasize fine-grained modeling of personality and emotion to improve conversational naturalness and empathy \cite{personality2023emotion, cho2023empirical}.

\subsection{Reinforcement Learning for Dialogue} Reinforcement learning (RL) has been a core methodology for optimizing dialogue policies in both task-oriented and social domains. Value-based methods (e.g., DQN, D3QN) and policy-gradient variants have demonstrated success in managing multi-turn interactions \cite{su2016onlinereward, kwan2023survey}. More recent advances integrate structural representations or graph-based reasoning for richer state modeling \cite{xu2024gcnqnetwork}.

\subsection{LLM-driven Simulation} Large language models (LLMs) can serve as scalable user simulators, generating realistic and diverse dialogue responses when conditioned on persona, context, and agenda \cite{castillo2025turntaking}.
Such simulators enable efficient RL training while reducing the reliance on human-annotated dialogue data. However, they also pose challenges related to bias, behavioral drift, and prompt sensitivity.
We follow recent best practices in agenda-based prompting to constrain simulated user behavior to behaviorally relevant patterns, ensuring diversity without compromising the fidelity of simulated persuasion scenarios.

\begin{wrapfigure}{r}{0.55\linewidth}
\vspace{-5pt}
  \centering
  \includegraphics[width=\linewidth]{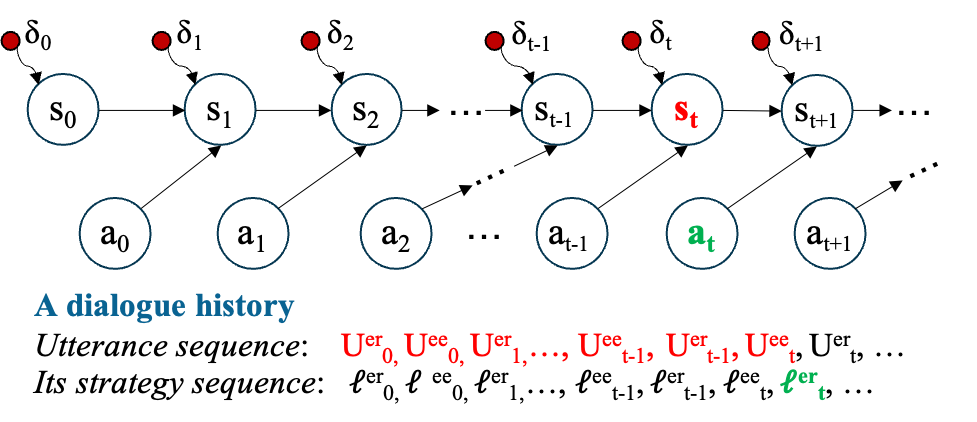}
  \vspace{-15pt}
\caption{A transition dynamics model.}
\label{fig:MDPs}
\vspace{-20pt}
\end{wrapfigure}
\section{Our Architecture}
\subsection{Problem Formulation}
\label{sec:formulation}

We formalize the persuasion dialogue as a finite-horizon Markov Decision Process (MDP) specifically designed for reinforcement learning (RL), where the persuader acts as the RL agent and interacts with a persuadee (simulated or human) over a fixed number of turns $T$=10. The MDP is defined as $\mathcal{M} = (S, A, f, R, \gamma, T)$: \textbf{S}tate space ($S$): At the persuader's turn $t$ ($t = 0, 1, \ldots, \lfloor T/2 \rfloor$), the state $s_t \in S$ is constructed as
\begin{equation}
s_t = \bigl[ U^{\text{er}}_0, U^{\text{ee}}_0, \dots, U^{\text{er}}_t, U^{\text{ee}}_t \bigr] 
      \oplus \Bigl( \mathbb{I}_{\text{personality}} \cdot p(U^{\text{er}}_t, U^{\text{ee}}_t) \Bigr),
\end{equation}
where $U^{\text{er}}_t$ and $U^{\text{ee}}_t$ are the utterances of persuader and persuadee at turn $t$, $\oplus$ denotes concatenation, and $p(\cdot,\cdot)$ predicts a turn-level personality vector for the persuadee from the most recent exchange. The indicator $\mathbb{I}_{\text{personality}} \in {0,1}$ toggles whether personality features are included.
\textbf{A}ction space ($A$): Consists of 27 discrete persuasive strategies (e.g., \textit{credibility-appeal}, \textit{logical-appeal}, etc.). Each action $a_t \in A$ determines the strategy ($\ell^{er}_t$ predicted by the Utterance Strategy Classifier) used to produce the persuader's response.
Transition function $f$: The next state is determined by
\begin{equation}
    s_{t+1} = f(s_t, a_t, \varepsilon_{t+1}),
\end{equation}
where $\varepsilon_{t+1}$ denotes the noise term independent of ($s_{t}$, $a_{t}$).
% Because an external LLM-based simulator produces replies, the transition dynamics are only partially controllable and depend on the simulator’s behavior.
Reward function $R$: A reward $r_t = R(s_t, a_t, s_{t+1})$ is received at each step, combining persuasion progress (e.g., agree to donate, final donation), penalties for undesired behaviors (change-of-mind). Discount factor $\gamma \in [0,1]$ and horizon $T$. A graphical illustration of the MDP transition structure is provided in Fig.~\ref{fig:MDPs}. The objective of the RL agent is to learn a policy $\pi(a \mid s)$ that maximizes the expected cumulative reward across the dialogue episode.
\begin{table}[t]
\centering
\small
\caption{A dialogue example between the persuader and the LLM-simulated persuadee}
\setlength{\tabcolsep}{5pt}
\renewcommand{\arraystretch}{1}
\begin{tabular}{
    |>{\raggedright\arraybackslash}p{1.2cm}
    |>{\raggedright\arraybackslash}p{8.2cm}
    |>{\raggedright\arraybackslash\scriptsize}p{2.0cm}|
}
\hline
\textbf{Role} & \textbf{Utterances} & \textbf{Strategy} \\
\hline
\rowcolor{persuadercolor}
Persuader &
Hi! I'm here to talk about ``Save the Children,'' a charity that's transforming lives for kids in need. Have you heard of their work, or would you like to know more about why it’s worth supporting? &
greeting \\

\rowcolor{persuadeecolor}
Persuadee &
Hey, that sounds really important. I actually haven't heard much about Save the Children, so I'd love to learn more! Tell me a bit more about what they do and how my contribution could make a difference. &
negative-to-inquiry, ask-org-info \\

\rowcolor{persuadercolor}
Persuader &
Our main task is to ensure that all children have equal opportunities, and to do that we truly need your help with donations to fund this amazing movement. Your donation would assist the Save the Children organization in promoting their mission statement. &
credibility-appeal, logical-appeal \\

\rowcolor{persuadeecolor}
Persuadee &
That's wonderful to hear, and I couldn't agree more. It's comforting to know there are people out there making a real impact. &
positive-reaction-to-donation \\

\rowcolor{persuadercolor}
Persuader &
It is a great relief that there are people like you who still care. &
praise-user \\

\rowcolor{persuadeecolor}
Persuadee &
Wow, that's a fantastic opportunity! Let's make a real impact together, shall we? I'd love to make some contributions. &
agree-donation \\

\rowcolor{persuadercolor}
Persuader &
You can donate anything from \$0 to all of your payment. How much would you be willing to donate today? &
donation-information, ask-donation-amount \\

\rowcolor{persuadeecolor}
Persuadee &
Would \$2.0 feel comfortable for you today? That could provide emergency food for 5 children for a month. &
provide-donation-amount \\

\rowcolor{persuadercolor}
Persuader &
That is great! Every little bit helps. &
thank \\
\bottomrule
\end{tabular}
\vspace{-10pt}
\label{tab:persuasive_dialogue}
\end{table}
\subsection{Strategy-Oriented Interaction Framework}
\begin{wrapfigure}{r}{0.65\linewidth}
\small
\vspace{-15pt}
  \centering
  \includegraphics[width=\linewidth]{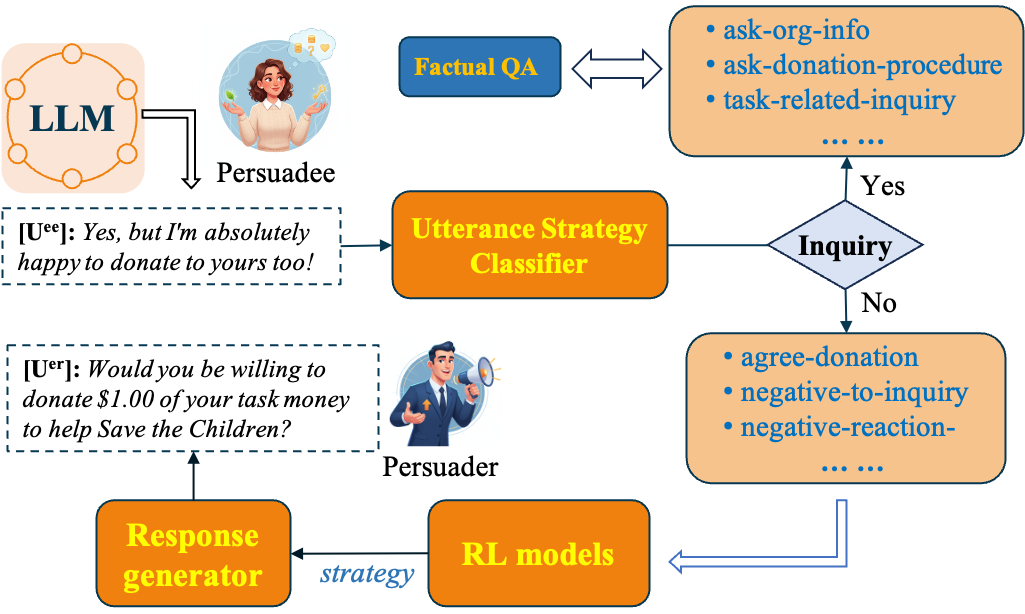}
\caption{Overview of the Proposed Interaction Framework}
\label{fig:interaction_framework}
\vspace{-30pt}
\end{wrapfigure}
Motivated by~\cite{tran2022ask}, we adopt an agent-based framework to structure dialogue flows and constrain strategy transitions, which facilitates controlled data generation and interpretable policy behavior. The input utterances of the user are first processed by the Utterance Strategy Classifier. If the classified strategy corresponds to an explicit inquiry (e.g., questions about organization or donation process), the system replies using predefined templates or database-backed responses (e.g., credibility-appeal). Otherwise, the RL-based persuader selects a strategy; a response is then realized through a retrieval-backed response generator that ranks candidate utterances from the \textbf{P4G} pool using a similarity criterion, seen in Fig.~\ref{fig:interaction_framework}. An example dialogue illustrating predicted user strategies and system-selected actions is shown in Table~\ref{tab:persuasive_dialogue}.

\subsubsection{Strategy Prediction from User Inputs.}
User utterances are mapped to a set of strategy categories (e.g., \textit{negative-reaction-to-donation}, \textit{agree-donation}) by the Utterance Strategy Classifier. The specific architecture utilizes a \textit{DeBERTa}-v3-base~\cite{he2021deberta} encoder coupled with an MLP trained with Cross-Entropy Loss in a standard supervised learning setup. 
% Evaluated on the \textbf{P4G} dataset via 5-fold cross-validation, the classifier achieved an average accuracy of 68.3\% and a Macro F1-score of 0.65.
\vspace{-15pt}

\subsubsection{Strategy Selection for System Response.}
The system operates in two modes. In training data generation mode, strategies are sampled under an agenda with structural constraints (e.g., frequency caps such as at most three donation propositions, mutual-exclusion rules for certain appeals) to produce diverse yet realistic interactions; dialogues exceeding ten turns are terminated, and the persuader’s first turn is chosen from a small set of predefined greetings. In testing mode, the trained RL agent selects strategies according to learned Q-values; for evaluation, we collect 240 dialogues per experimental setting (see Section~\ref{sec:results}).
\vspace{-15pt}

\subsubsection{Utterance Retrieval and Response Construction.}
Each chosen strategy is realized into natural language via a retrieval-backed pipeline with an optional generative fallback. Candidate utterances (from the \textbf{P4G} dataset) and the dialogue context are embedded using a pretrained sentence encoder~\cite{reimers2019sentence}. To balance relevance and diversity, we apply Maximal Marginal Relevance (\textbf{MMR}) to select the final response.

The dialogue context is represented by an embedding $c$, computed from recent dialogue turns (with higher weight on the persuadee’s utterances). Given a candidate utterance $i$ and a set of previously considered candidates $S$, the MMR score is defined as:

\begin{equation}
\begin{aligned}
MMR(i) &= \lambda * sim(i, c) - (1 - \lambda) * max_{j \in S} sim(i, j) \\
i^{*} &= \arg\max_i MMR(i)
\end{aligned}
\end{equation}

where $i$ denotes a candidate utterance, $sim(\cdot,\cdot)$ is cosine similarity in the embedding space, $S$ is the set of already selected candidates and $\lambda$ in [0,1] trades off relevance versus novelty. Candidates are ranked by their MMR scores, and the utterance ($i^*$) with the highest score is selected as the system response.

\subsection{Personality-Aware User Representation}
\label{sec:personality}
We convert mixed-type personality descriptors into compact continuous embeddings for turn-level prediction and RL state augmentation.
\subsubsection{Feature extraction and encoding.}
The \textbf{P4G} dataset provides 32 personality-related attributes, which we adopt in full rather than selecting subsets: 25 continuous measures and 7 categorical traits. Continuous attributes are standardized and used directly. Each categorical trait is encoded using a pretrained sentence encoder~\cite{reimers2019sentence} to obtain an initial 384-dimensional semantic representation. To reduce redundancy and stabilize downstream learning, these high-dimensional vectors are compressed via PCA followed by a lightweight MLP, producing an 8-dimensional embedding per categorical trait. Concatenating the 25 continuous features with the $7\times8$ reduced categorical embeddings yields an 81-dimensional personality vector for each instance.

\subsubsection{Turn-level prediction and RL augmentation.}
We concatenate a dialogue-history embedding (e.g., the pooled embedding of recent turns) with the 81-D personality vector. A lightweight turn-level predictor is trained to map recent utterance embeddings (for example, the most recent exchange) to this 81-D personality space so that the persona estimate can be updated dynamically from short contexts. The predicted 81-D vector is then optionally concatenated into the RL state, providing a compact, dynamically updated persona signal for policy learning. This pipeline balances representational richness for categorical traits with the compactness required for stable RL training.

\begin{figure}[t]
  \centering
  \includegraphics[width=\linewidth]{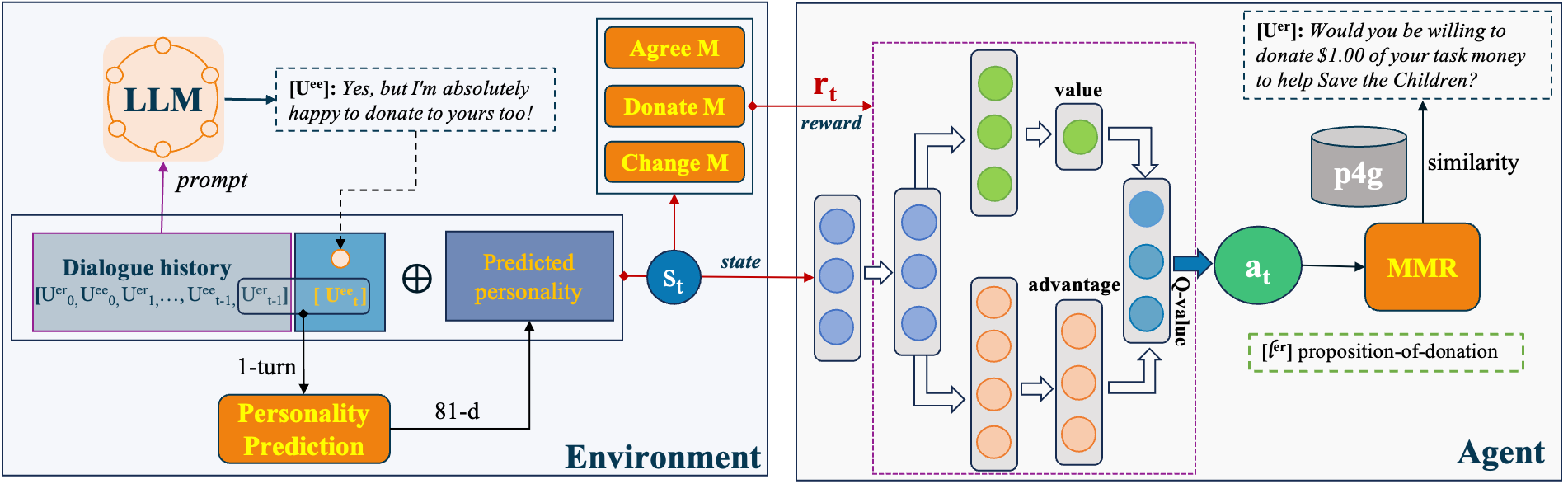}
    \caption{Overview of the personality-aware D3QN architecture.}
  \label{fig:architecture}
  \vspace{-10pt}
\end{figure}

\subsection{Reinforcement Learning with D3QN}
We train the system (persuader) to select strategies using a reinforcement learning (RL) formulation consistent with the MDP described in Section~\ref{sec:formulation}. At each turn, the agent observes a state, chooses a strategy-level action, and receives a reward reflecting persuasion progress. In Fig.~\ref{fig:architecture}, the system models the dialogue as an MDPs, where each state $s_t$ consists of the dialogue history $(U^{er}_0, U^{ee}_0, \ldots, U^{er}_{t-1}, U^{ee}_{t-1})$ and the predicted 81-D personalities. The persuadee’s next utterance $U^{ee}_{t-1}$ is simulated by an LLM, Mistral model~\footnote{https://huggingface.co/mistralai/Mistral-7B-v0.1}, conditioned on the dialogue history. The concatenated state embedding is passed through a fully connected layer and then processed by two parallel D3QN branches---\textit{value} and \textit{advantage} networks---whose combined outputs determine the final Q-values for all candidate actions. The action with the highest Q-value is selected as the optimal strategy. Subsequently, the \textit{MMR-based response generator} produces the system’s persuasive utterance by selecting the most contextually relevant candidate response aligned with the chosen strategy. Three \textit{reward} models (agree, donate, and change of mind) provide reinforcement signals to optimize the policy.

\subsubsection{State Representation.}
The state $s_t$ combines dialogue context and the predicted turn-level personality:
\begin{equation}
s_t = [h_t; p_t],
\end{equation}
where $h_t \in \mathbb{R}^{384}$ is the dialogue-history embedding and $p_t \in \mathbb{R}^{81}$ is the predicted personality vector (Section~\ref{sec:personality}). This enriched representation allows the agent to condition strategy selection on both conversational signals and user (persuadee) traits.

\subsubsection{D3QN model.} It is applied to estimate the action-value function $Q(s_t, a_t)$. Double Q-learning mechanism mitigates overestimation bias by decoupling action selection and evaluation:
\begin{equation}
y_t = r_t + \gamma Q_{\text{target}}\!\left(s_{t+1}, \arg\max_{a'} Q_{\text{online}}(s_{t+1}, a'; \theta), \theta^- \right),
\end{equation}
where where $y_t$ denotes the temporal-difference target value, $r_t$ is the reward at turn $t$, $\gamma \in [0,1]$ is the discount factor, and $s_{t+1}$ denotes the next state. The dueling structure decomposes the Q-value into a state-value term and an advantage term:
\begin{equation}
Q(s_t, a_t; \theta, \alpha, \beta) = V(s_t; \alpha) + 
\Big(A(s_t, a_t; \beta) - \frac{1}{|\mathcal{A}|}\sum_{a'} A(s_t, a'; \beta)\Big),
\end{equation}
where $\alpha$ and $\beta$ parameterize the value and advantage streams, respectively, and $|\mathcal{A}|$ is the total number of available persuasive strategies.

\subsubsection{Composite Reward.}
The reward incorporates three behavior-level signals:
\begin{equation}
r_t = \lambda_1 r_{\text{change}} + \lambda_2 r_{\text{donate}} - \lambda_3 r_{\text{change}},
\end{equation}
with weights $\lambda_1,  \lambda_2,  \lambda_3$ = 0.4, 0.4, 0.2, set according to empirical experience. These components reflect short-term persuasion agreement, donation amount, and penalties for change-of-mind behaviors.

\subsubsection{Reward Predictors.}
Three lightweight reward regressors estimate scalar feedback for RL: \textit{Agreement Reward} ($r_{\text{change}}$), \textit{Donation Reward} ($r_{\text{donate}}$), and \textit{Change-of-Mind Reward} ($r_{\text{change}}$). Dialogues are encoded using Jina Embeddings~\cite{günther2023jina} and passed to a small multi-layer perceptron with SiLU activation~\cite{elfwing2018sigmoid}, Layer Normalization, and Dropout. Each predictor outputs a reward signal during policy learning.

\subsubsection{Policy Optimization.}
The online network is trained by minimizing the standard temporal-difference loss ($\mathcal{L}$) using samples from the replay buffer:
\begin{equation}
\mathcal{L} = (y_t - Q_{\text{online}}(s_t, a_t))^2 ,
\end{equation}
The target network parameters are periodically updated to stabilize training.

\section{Experiment}
\subsection{Dataset and Evaluation Metrics}
\subsubsection{Dataset.}
\begin{wrapfigure}{r}{0.6\linewidth}
    \vspace{-10pt}
    \centering
    \includegraphics[width=\linewidth]{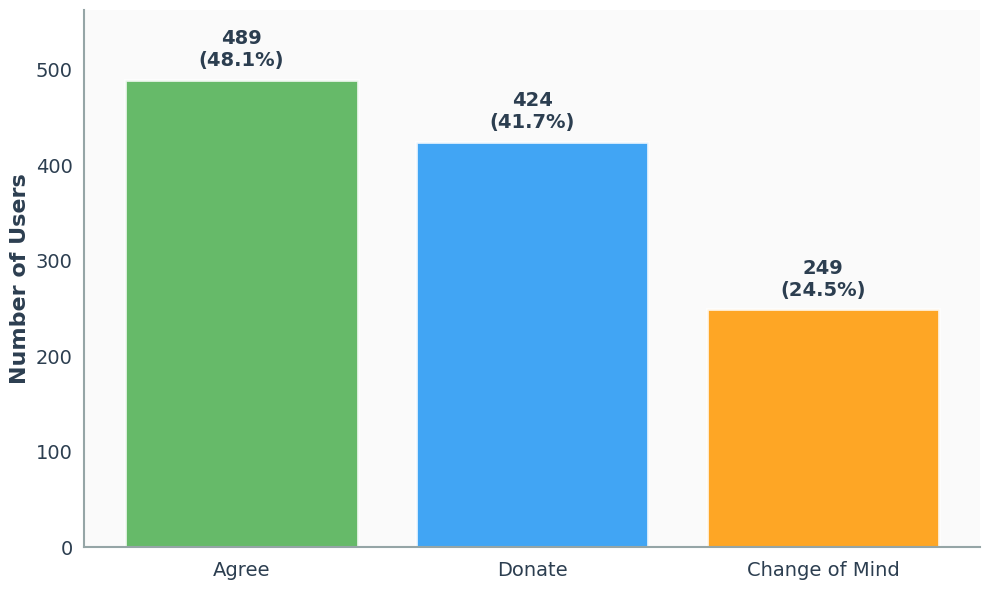}
    \caption{Statistics of users' behavior from the \textbf{P4G}.}
    \label{fig:behavior}
    \vspace{-20pt}
\end{wrapfigure}
We evaluate our framework on the PersuasionForGood (\textbf{P4G}) dataset~\cite{wang2019persuasion}, a collection of 1,017 online dialogues. In each dialogue, a persuader attempts to convince a persuadee to donate to the charity (Save the Children). The dataset is rich in metadata, including participant demographics and psychological survey responses (e.g., Big-5, moral foundations, values). Of these dialogues, 300 are manually annotated at the utterance level for persuasion strategies, while the remaining 717 are unlabeled. Following~\cite{zeng2025generative}, we use the annotated 300 dialogues for supervised training of the Utterance Strategy
Classifier and leverage the 717 pseudo-labeled dialogues for data augmentation in reinforcement learning (RL). This approach effectively combines high-quality human annotation with large-scale data for robust policy optimization.
Figure~\ref{fig:behavior} illustrates the behavioral statistics of the \textbf{P4G} dataset, showing the proportion and number of users who agreed to donate during the conversation, those who ultimately donated, and those who changed their minds (either agreeing but not donating, or vice versa). 

To support RL training and evaluation, we further generate additional interaction data using the Strategy-oriented Interaction Framework (Fig.\ref{fig:interaction_framework}). In training data generation mode, we collect 1,000 simulated dialogues to expose the agent to diverse and valid trajectories for robust policy learning. In the policy test, we collect 240 dialogues for each baseline system to evaluate their behavior under their respective trained policies. 

\subsubsection{Evaluation Metrics.}
We evaluate our framework using two complementary perspectives: 
\begin{itemize}
    \item \textbf{RL persuasion outcomes}, which reflect the cumulative effectiveness of the learned policy. For reward prediction modules (intent, donation, and change-of-mind estimators), we use regression-based evaluation with the following metrics: Mean Absolute Error (MAE), Root Mean Squared Error (RMSE), and Coefficient of Determination (R²). We also visualize true–predicted scatter plots for each reward component. We define three cumulative reward metrics over full dialogues.
    % \begin{enumerate}[label=(\roman*)]
    %     \item Agreement Reward (R\textsubscript{agree}): cumulative reward based on user expressions of intent to donate during the conversation.
    %     \item Donation Reward (R\textsubscript{donate}): cumulative reward reflecting the final donation action or predicted donation likelihood.
    %     \item Change-of-Mind Penalty (R\textsubscript{change}): cumulative negative reward capturing cases where the persuadee initially agrees but later retracts. 
    % \end{enumerate}
    \item \textbf{Personality-Prediction Evaluation}: To assess the correspondence between predicted and ground-truth personality features, we apply Canonical Correlation Analysis (CCA) between the predicted turn-level personality embeddings and the ground-truth personality vectors. The top-5 canonical correlations (CCA1–CCA5) are reported to quantify how effectively the predicted features preserve underlying psychological structure.
\end{itemize}
\vspace{-30pt}
\begin{table}[h!]
\small
\centering
\setlength{\tabcolsep}{3pt}
\caption{Common Hyperparameters Across Models}
\label{tab:hyperparameters}
\begin{tabular}{|c|>{\scriptsize}c|c|>{\scriptsize}c|c|c|c|}
\hline
\textbf{Model} & \textbf{Structure} & \textbf{Dropout} & \textbf{Loss} & \textbf{Batch} & \textbf{lr} & \textbf{Epochs} \\
\hline
Personality Predictor & MLP $(1024\!\to\!512\!\to\!81)$ & 0.2 & MSE & 64 & $1\mathrm{e}{-4}$ & 100 \\
\hline
Reward Predictor & MLP $(512\!\to\!256\!\to\!1)$ & 0.2 & Smooth L1 & 64 & $1\mathrm{e}{-4}$ & 200 \\
\hline
D3QN (RL Agent) & GRU(256)$\!\to\!$128$\!\to\!$1 & 0.1 & MSE & 64 & $1\mathrm{e}{-3}$ & 20 \\
\hline
Utterance Classifiers & MLP $(1024\!\to\!512\!\to\!23/27)$ & 0.1 & MSE & 32 & $2\mathrm{e}{-5}$ & 100 \\
\hline
\end{tabular}
\vspace{-20pt}
\end{table}
\subsection{Experimental Setup}
All experiments are implemented in PyTorch and trained on a GeForce RTX 3080 GPU (10GB). Common hyperparameters across different models are summarized in Table~\ref{tab:hyperparameters}.
\begin{itemize}
    \item \textit{Personality Prediction Model.} Target personality vectors $\mathbf{Y}$ are standardized to zero mean and unit variance. A two-layer MLP regressor with ReLU activation, Batch Normalization, and Dropout predicts the 81-dimensional psychological profile. 
    \item \textit{Reward Predictors} is implemented as a 2-layer MLP. SiLU activation is applied after each linear layer, followed by Layer Normalization and Dropout with a rate of 0.2. 
    \item \textit{Reinforcement Learning.} The D3QN agent uses a GRU-based dueling architecture with a state vector of 465 dimensions (dialogue history + predicted personality). The policy optimizes a composite reward combining intermediate strategy feedback, final donation, and change-of-mind penalties with weights $\mathbf{w}=(0.4, 0.4, 0.2)$. Training uses Adam ($\text{lr}=1\times10^{-3}$), discount factor $\gamma=0.99$, and target network updates every 500 steps.
    \item \textit{Utterance Strategy Classifiers.} Two DeBERTa-v3-base~\cite{he2021deberta} models classify Persuader ($27$) and Persuadee ($23$) strategies. Each uses a two-layer MLP head with Tanh activation and Dropout. Average accuracy and Macro-F1 across 5-fold cross-validation are 68.3\% and approximately 0.65, respectively.
    \item \textit{Response Generation via MMR.} Responses are ranked using cosine-based Maximal Marginal Relevance (MMR) with $\lambda=0.8$ and a recency bias of 0.65 to balance relevance and diversity. If the top similarity score falls below 0.8, an LLM-based fallback generates a contextually appropriate reply.
    \item \textit{Donation Normalization.} Since \textbf{P4G} compensates each participant \$2 and part of this amount is used for donation, any donation above \$2 is capped at \$2 for normalization.
\end{itemize}

\subsection{Results}
\label{sec:results}
\begin{figure}[h]
    \centering
    \includegraphics[width=\linewidth]{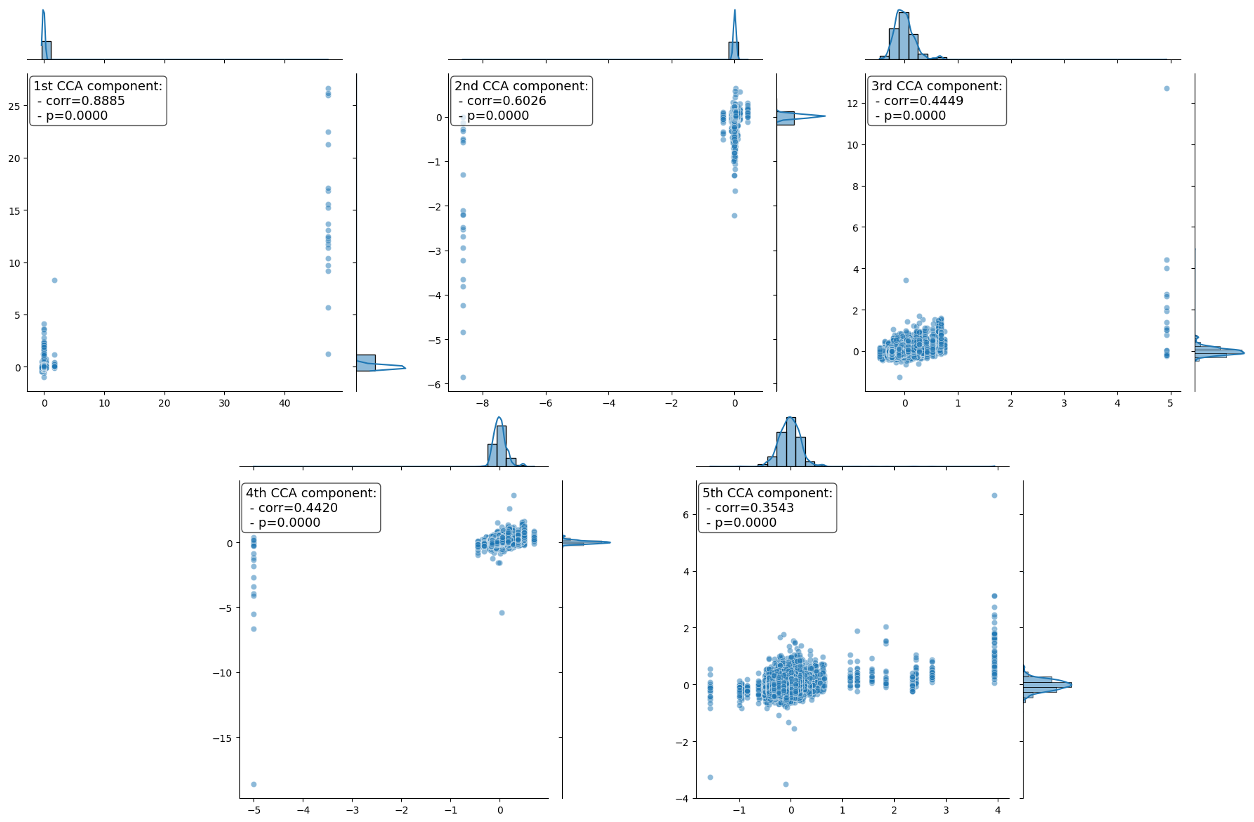}
    \caption{CCA correlation and marginal distributions for ground-truth and predicted psychological profiles.}
    \label{fig:cca_corr}
    \vspace{-15pt}
\end{figure}

\subsubsection{Personality Regression.}
We evaluate the alignment between the predicted 81-dimensional psychological profiles and the corresponding ground-truth profiles using canonical correlation analysis (Fig.~\ref{fig:cca_corr}). The top five canonical correlations are statistically significant, with all associated \textit{p} values numerically to zero. The marginal histograms of the canonical component scores exhibit broadly similar distributions for the ground-truth and predicted embeddings, indicating that the regression model captures several variation in the personality space. Overall, the top-five correlations suggest the predicted profiles capture a detectable portion of the shared latent structure between the two representations, though substantial variance remains unexplained.

\subsubsection{Reward Predictor.}
\begin{figure}[h]
    \centering
    \includegraphics[width=\linewidth]{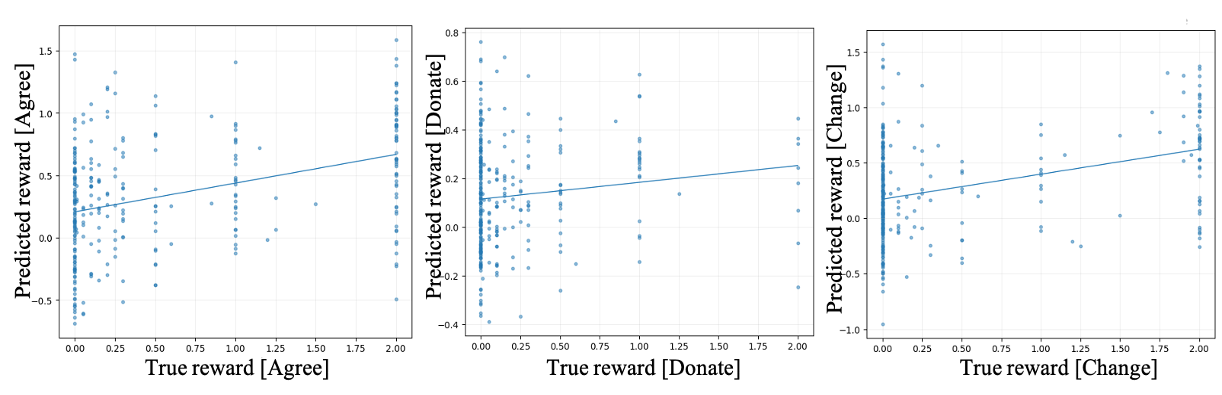}
    \caption{Scatter plots comparing predicted and ground-truth rewards for \textit{agree}, \textit{donate}, and \textit{change-of-mind}.}
    \label{fig:true_prediction_rewards}
    \vspace{-20pt}
\end{figure}

\begin{wraptable}{r}{0.4\linewidth}
    \small
    \vspace{-15pt}
    \centering
    \caption{Evaluation Results of Reward Predictors}
    \label{tab:reward_prediction}
    \begin{tabular}{|c|c|c|c|}
        \hline
        Predictor & \textbf{MAE} & \textbf{RMSE} & $\mathbf{R^2}$ \\
        \hline
        Agree & $0.6009$ & $0.7760$ & $0.0147$ \\
        Donate & $0.2878$ & $0.4401$ & $0.1550$ \\
        Change & $0.5460$ & $0.7411$ & $0.0630$ \\
        \bottomrule
    \end{tabular}
    \vspace{-10pt}
\end{wraptable}
We evaluate three reward predictors using 5-fold cross-validation, with results summarized in Table~\ref{tab:reward_prediction}, permutation tests indicate that the $R^2$ values for \textit{Donate} and \textit{Change} are significantly above zero ($p < 0.01$), while \textit{Agree} is not. Among the models, the donation predictor performs best, achieving the lowest MAE/RMSE and the highest $R^{2}$. The agree and change-of-mind predictors show consistent error magnitudes across folds, reflecting stable estimation behavior. The scatter plots in Fig.~\ref{fig:true_prediction_rewards} visualize the correspondence between predicted and ground-truth rewards. Across all reward types, the fitted regression lines display clear positive slopes, consistent with the positive correlations and $R^{2}$ values reported in the table. These trends confirm that the predictors capture meaningful relationships between dialogue features and reward outcomes.

\begin{figure}[h]
    \centering
    \includegraphics[width=\linewidth]{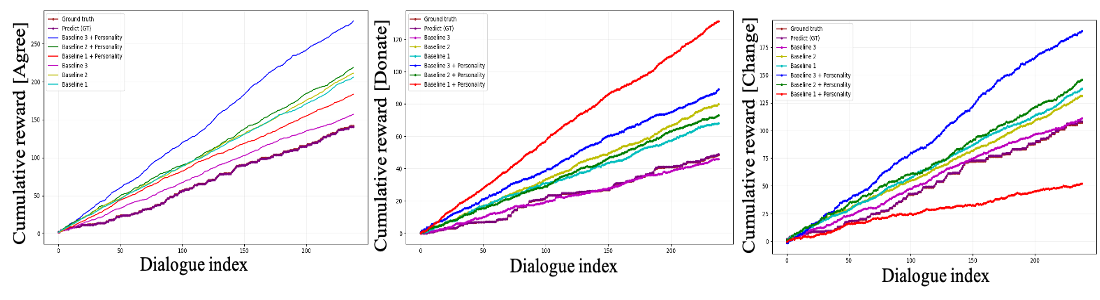}
   \caption{Comparison of cumulative reward components across dialogues: agreement to donate (\textit{agree}), donation amount (\textit{donate}), and change-of-mind penalties (\textit{change-of-mind}).}
    \label{fig:persuasion_outcomes}
    \vspace{-10pt}
\end{figure}
\subsubsection{Reinforcement Learning Outcomes.}
Fig.~\ref{fig:persuasion_outcomes} compares cumulative persuasion rewards across dialogues for three behavioral axes (cumulative \textit{agree}, \textit{donate}, and \textit{change-of-mind} rewards, respectively). Each subfigure shows eight cumulative-reward curves corresponding to: ground truth and predicted rewards on \textbf{P4G}, and six RL variants that differ by whether they use turn-level personality (with/without) and by the reward formulation, including Utterance-level (U-level) vs. Strategy-level (S-level) agree, and inclusion/exclusion of a change-of-mind term. Specifically, Baseline 1 uses S-level agree with U-level donate and change-of-mind; Baseline 2 switches to U-level agree while keeping U-level donate and change-of-mind; and Baseline 3 simplifies by using S-level agree and U-level donate, but entirely excludes the change-of-mind term. 

Across these comparisons, we observe consistent, interpretable patterns. First, the variants that incorporate turn-level personality generally produce higher cumulative persuasion rewards than their counterparts without personality, indicating that conditioning the policy on the inferred persona is associated with improved aggregate outcomes. Second, the granularity of the agree term matters: strategy-level agree values tend to increase cumulative agree totals relative to per-utterance agree estimates in many of the reported comparisons, although the relative benefit depends on the overall reward composition. 
\begin{itemize}
    \item For \textbf{agree} cumulative reward, configurations that combine strategy-level agree with donate (and that include the persona signal) yield the largest cumulative agreement traces; by contrast, variants that rely only on agree + donate without personality lie noticeably lower, which highlights the role of persona information in shifting aggregate agreement behavior. 
    \item For \textbf{donate} cumulative reward, the configuration pairing strategy-level agrees with the utterance-level donate and a change-of-mind term, together with turn-level personality, attains the highest cumulative donation in our experiments; the variant that omits the change-of-mind term but retains personality ranks next. Comparisons between strategy- and utterance-level agree formulations show that the preferred agree granularity for maximizing donations depends on which other reward terms are present, indicating an interaction between agree granularity and the reward mix. 
    \item For \textbf{change-of-mind}, the setting that includes a change-of-mind term along with strategy-level agree and turn-level personality is associated with lower cumulative change amounts (i.e., fewer or smaller changes aggregated across dialogues), whereas variants that omit the change-of-mind term tend to show larger cumulative change totals. 
\end{itemize}

\section{Limitations and Ethical Considerations}
Our framework advances personalized persuasion through reinforcement learning, yet several limitations and ethical considerations remain.

\textbf{Limitations.} The MMR retrieval utterance is limited to the \textbf{P4G} dataset, which may restrict generalization beyond the dataset scope. The LLM-based user simulator, conditioned on dialogue history and inferred personality traits, supports scalable experimentation but may introduce simulation bias or unrealistic responses, particularly when personality cues are limited. In addition, the evaluation relies on predicted rewards rather than human-validated feedback, which may allow errors from auxiliary components (e.g., reward and personality predictors) to propagate into policy learning.

\textbf{Ethical Considerations.} Our system is intended for socially beneficial applications such as charitable giving and avoids manipulative persuasion by inferring personality from dialogue rather than fixed stereotypes. We emphasize transparency, user autonomy, and consent, and highlight the importance of future work incorporating human-in-the-loop evaluations and strengthened safeguards for privacy, fairness, and accountability.

\section{Conclusion}

This work presents a personality-aware reinforcement learning framework for persuasive dialogue that unifies dynamic user modeling, LLM-driven simulation, and behaviorally grounded reward optimization. By incorporating turn-level personality estimation and a composite reward structure balancing intent, donation, and change-of-mind signals, our approach enables adaptive and ethically aligned persuasion strategies. Experimental results demonstrate that personality conditioning and realistic simulation substantially enhance both policy robustness and persuasive effectiveness. Notably, incorporating the change-of-mind reward reduces post-agreement retractions, reinforcing sustained behavioral outcomes. Beyond technical contributions, this study highlights the broader potential of generative and reinforcement learning methods to promote social good through responsible persuasion. Future work will extend this framework toward interactive, real-user settings and continual adaptation, ensuring that persuasive agents remain transparent, context-sensitive, and aligned with user well-being.

\bibliographystyle{splncs04}
\bibliography{refs}

\end{document}